\documentclass{article}
\usepackage{authblk}
\usepackage{graphicx}
\usepackage{xspace}
\usepackage{subcaption}
\usepackage{hyperref}
\newcommand{\etal}{et~al.\xspace}

\newcommand{\eg}{{e.g.,\@ }}
\newcommand{\vfour}{IPv4\xspace}
\newcommand{\vsix}{IPv6\xspace}
\newcommand{\eui}{EUI-64\xspace}
\newcommand{\bap}{byte-axis plot\xspace}
\newcommand{\wifi}{Wi-Fi\xspace}

\title{Visualizing MAC and IPv6 Address Allocations}
\author{Erik Rye}
\affil{Computer Science Department\\University of Maryland\\
rye@cs.umd.edu}

\begin{document}
\maketitle

\section{Introduction \& Background}

In this work, I describe a method for visualizing two types of network address
allocations: Media Access Control (MAC) addresses --- hardware identifiers
permanently assigned to network interfaces --- and Internet Protocol Version 6
(\vsix) allocations --- subnetworks assigned from a larger, encompassing network
(\emph{e.g.,} allocated /64 networks from a /48 network).

It is often useful in practice to visualize the allocation of these two types of
addresses. Because the number of possible MAC addresses within an allocated
block (called an OUI, discussed \S\ref{sec:baps}) is large ($2^{24}$),
individual instances of observed MAC addresses provide
little in the way of context about how the OUI-owner
distributes them. In the \vsix case, individual active addresses or
active /64s subnetworks of a /48 or larger fail to supply useful information about the
allocation policies of the network operator. When individual
instances of MAC and IPv6 addresses are aggregated, however, they can often be
used to form
a helpful visualization of the assignment strategies in place
by various hardware manufacturers and network operators. As we shall see, both
MAC address and IPv6 subnet allocation strategies can vary wildly between
different hardware vendors and service providers. Reverse-engineering
these policies can be useful for applications that require, for instance,
understanding how many MAC addresses are assigned to individual devices
\cite{rye2022ipvseeyou} or the size of IPv6 subnetwork an ISP assigns to
customers~\cite{rye2021follow}.

There is some previous work in visualizing IP address allocations.
Dainotti~\etal created a $12^{th}$-order Hilbert map of the allocated \vfour
address space~\cite{dainotti2016lost}. In \vsix, Gasser~\etal
visualized /32 prefixes by clustering them according to the entropy of the
address nybbles, which they call their ``entropy
fingerprints''~\cite{Gasser:2018:CEU:3278532.3278564}. They also color and
visualize BGP prefixes by their ``cluster IDs,'' displaying common addressing
patterns among related /32 prefixes.

\begin{figure*}[t]
\centering
    \begin{subfigure}[c]{.48\textwidth}
        \includegraphics[width=\linewidth]{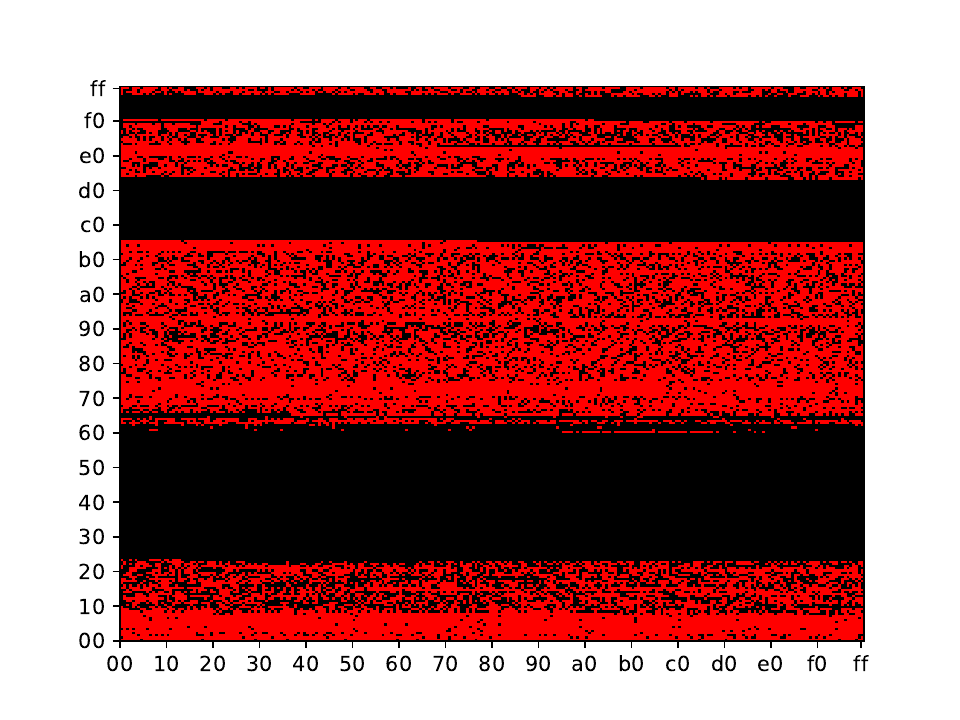}
        \caption{The byte-axis plot for the \texttt{08:3C:0C} (Arris) OUI.}
        \label{fig:arris-mac}
    \end{subfigure}%
    \hfill
    \begin{subfigure}[c]{.48\textwidth}
        \includegraphics[width=\linewidth]{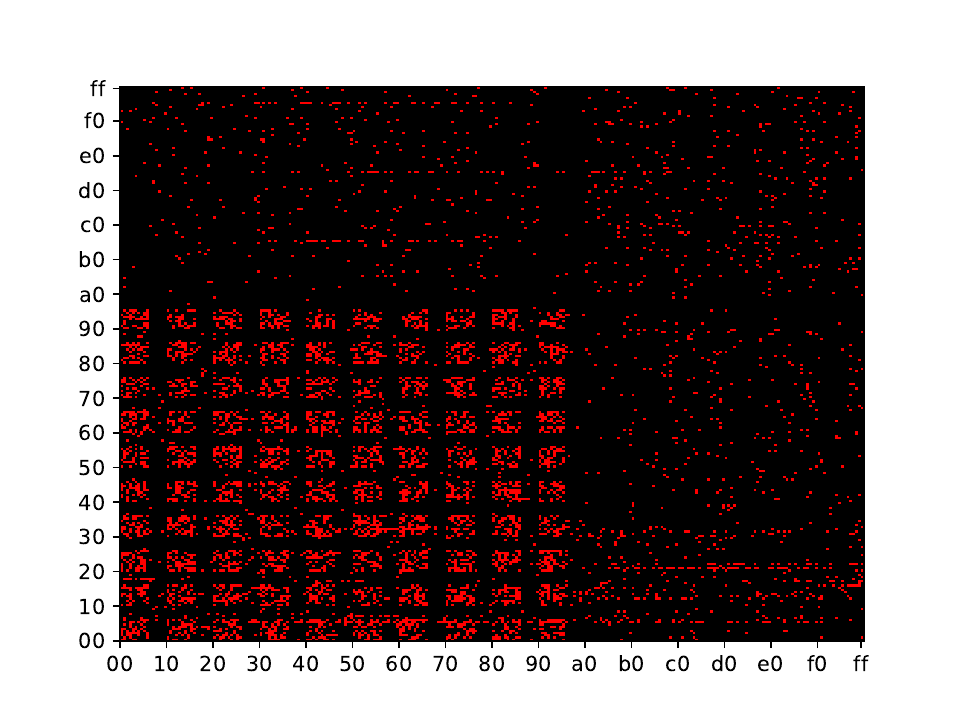}
        \caption{The byte-axis plot for the \texttt{00:27:15} (Rebound Telecom)
        OUI.}
        \label{fig:rebound-mac}
    \end{subfigure}%
    \caption{Two byte-axis plots for MAC address observations from a forthcoming
    study. Note the obvious variation in allocation strategies between the two
    manufacturers.}
    \label{fig:mac-plots}
\end{figure*}

In the next section, I describe my methodology for visualizing MAC and \vsix
address allocations within some larger super-allocation and provide practical
examples of its utility.

\section{Byte-Axis Plots}
\label{sec:baps}

I call the visual representations of the MAC and \vsix space described hereafter
``byte-axis plots.'' This name reflects the primary feature of these plots,
which is a separation of two of the low order bytes of these address types
along the $x$ and $y$ axes. While the MAC address and \vsix plots share many
similarities, there are enough subtle differences in what the two depict that I shall
describe them each separately in the following two subsections.

\subsection{MAC Address Allocation Byte-Axis Plots}

MAC addresses are 48-bit identifiers, typically written as six
hexadecimal values separated by colons or a dash, \eg
\texttt{00:11:22:33:44:55}. Ignoring the case of so-called ``locally-assigned''
MAC addresses, globally-unique MAC addresses are assigned from IEEE-allocated
blocks called Organizationally Unique Identifiers (OUIs). OUIs are the first
three bytes of the MAC address, \eg the \texttt{00:11:22} bytes in the previous
example\footnote{The IEEE ceased
using the term ``OUI'' in favor of three MAC address allocations of differing
sizes they call MAC Address - Large (MA-L), MAC Address - Medium (MA-M), and MAC
Address - Small (MA-S) allocations, which contain $2^{24}$, $2^{20}$, and
$2^{12}$ addresses, respectively. The MA-L allocation is equivalent to the
former OUI in all ways but name. Very few entities register
allocations in either of the smaller allocation sizes.  Due to the ubiquity of the
term OUI in the literature and practice, I continue to use it here.}. An
organization assigned an OUI by the IEEE can allocate the lower three
address bytes as it deems fit. MAC addresses are assigned to interfaces on a
device, \eg the wireless interface on a \wifi access point or Ethernet interface
on a desktop computer. Every 802.11 \wifi or Ethernet interface must have a MAC
address assigned before it can communicate on its local LAN. The relationship
between OUIs and organizations is many-to-one; large equipment vendors like
Cisco and Apple have several hundred OUIs registered\footnote{An up-to-date
listing of the OUI assignments from the IEEE is available at
\url{https://standards-oui.ieee.org/oui/oui.txt}}. 

I will use the two subplots of Figure~\ref{fig:mac-plots}
for MAC address \bap demonstration purposes. Every MAC address \bap is a
representation of the observed, allocated addresses from within a single OUI. Imagine
that I have observed a number of MAC addresses from the same OUI in use --
perhaps from observing the MAC addresses of clients on a LAN, or by extracting
them from \eui \vsix addresses. In order to detect whether there are
periodicities or patterns in the distributions of MAC addresses in this OUI, we
can use byte-axis plots as a visual aid. To begin, label the $x$ and $y$ axes
from 0-255; because MAC addresses are typically written in hexadecimal, tick
marks at multiples of 16 (\eg 0x00, 0x10, 0x20, $\dots$) are useful visual cues.
In my formulation, the $y$-axis has represented the fourth byte of MAC addresses
from within the \bap OUI, while the $x$ axis represents the fifth byte. A point
$(x, y)$, then, where $0 \leq x, y \leq 255$, represents 256 distinct MAC
addresses that share the same first five bytes -- \eg \texttt{00:11:22:y:x:z},
where $0 \leq z \leq 255$.

I initialize the \bap so that each $(x,y)$ point is some background color, which
is typically black or white. Then, for each MAC address present in the dataset to
visualize, I extract the fourth ($y$) and fifth ($x$) bytes, and make the cell
at $(x,y)$ some non-background color. If I am attempting to visualize the
overall allocation patterns within the OUI, a single color will suffice.
Figure~\ref{fig:arris-mac}, for example, represents an OUI assigned to the Arris Corporation,
a modem and wireless router manufacturer. Each red pixel in the \bap represents
at least one observation of an Arris MAC address, while black pixels indicate that no MAC
addresses, with the first five bytes \texttt{08:3C:0C:$y$:$x$}, where $x$ and $y$
are the $x$ and $y$ coordinates, were observed. The \bap depicts four clear bands,
each with a different size. The lowest band in the plot covers MAC addresses
ranging from \texttt{08:3C:0C:00:} -- \texttt{08:3C:0C:28:}; the largest band
covers MAC addresses from about \texttt{08:3C:0C:68:} -- \texttt{08:3C:0C:B8:}.
Two smaller bands are located in the upper half of Figure~\ref{fig:arris-mac}.

Figure~\ref{fig:rebound-mac} depicts an OUI assigned to Rebound
Telecommunications Company, and illustrates both the differences in allocation
schemes observed in the wild between device manufacturers, as well as highlights
the types of analysis enabled by using byte-axis plots. In this figure, a
clearly-defined red rectangle, itself comprised of smaller red allocated
rectangles, is immediately apparent. The presence of the smaller allocation blocks,
with interspersed columns and rows of unallocated black pixels, is indicative of
a deliberate allocation strategy that deviates substantially from that in the OUI
depicted in Figure~\ref{fig:arris-mac}. It is possible that the rows and columns
of black pixels represent MAC addresses not collected by our study (\eg Wi-Fi
MAC addresses for devices whose wired interfaces I capture and display in red.)
It is also possible that these MAC addresses are truly unallocated, and are
perhaps being held in reserve by the manufacturer for some later time.

\begin{figure*}[t]
\centering
    \begin{subfigure}[c]{.48\textwidth}
        \includegraphics[width=\linewidth]{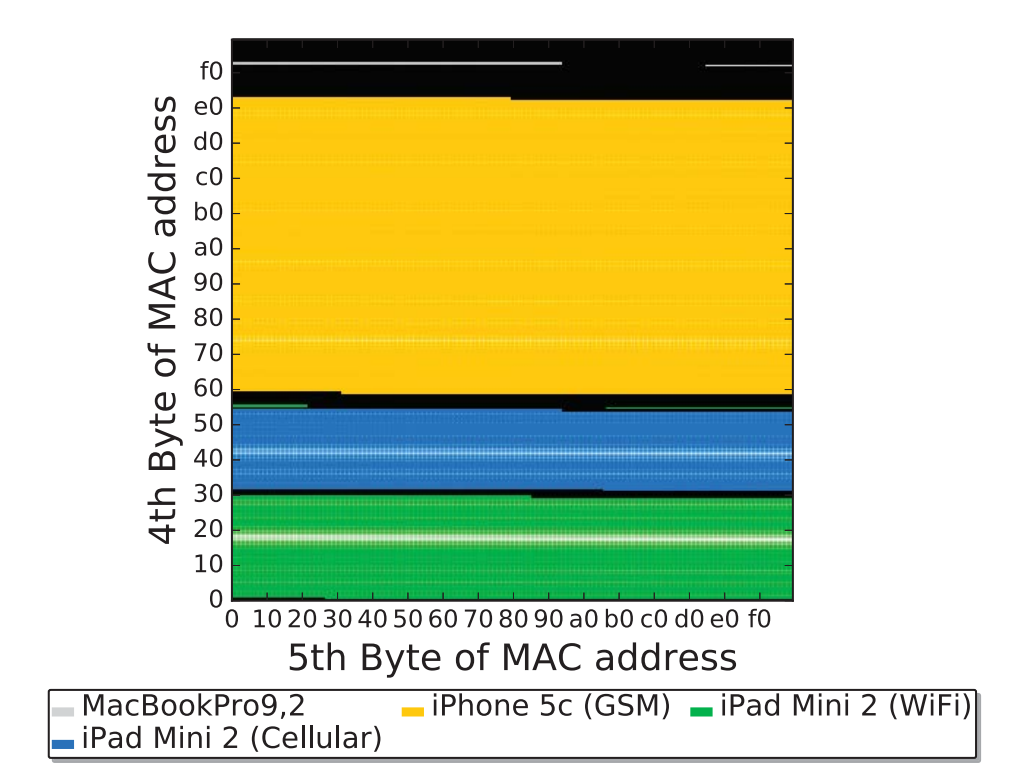}
        \caption{A visualization of an Apple OUI. Four different models are
        represented, although the number of MACs allocated to the four models
        varies significantly.}
        \label{fig:multi-apple}
    \end{subfigure}%
    \hfill
    \begin{subfigure}[c]{.48\textwidth}
        \includegraphics[width=\linewidth]{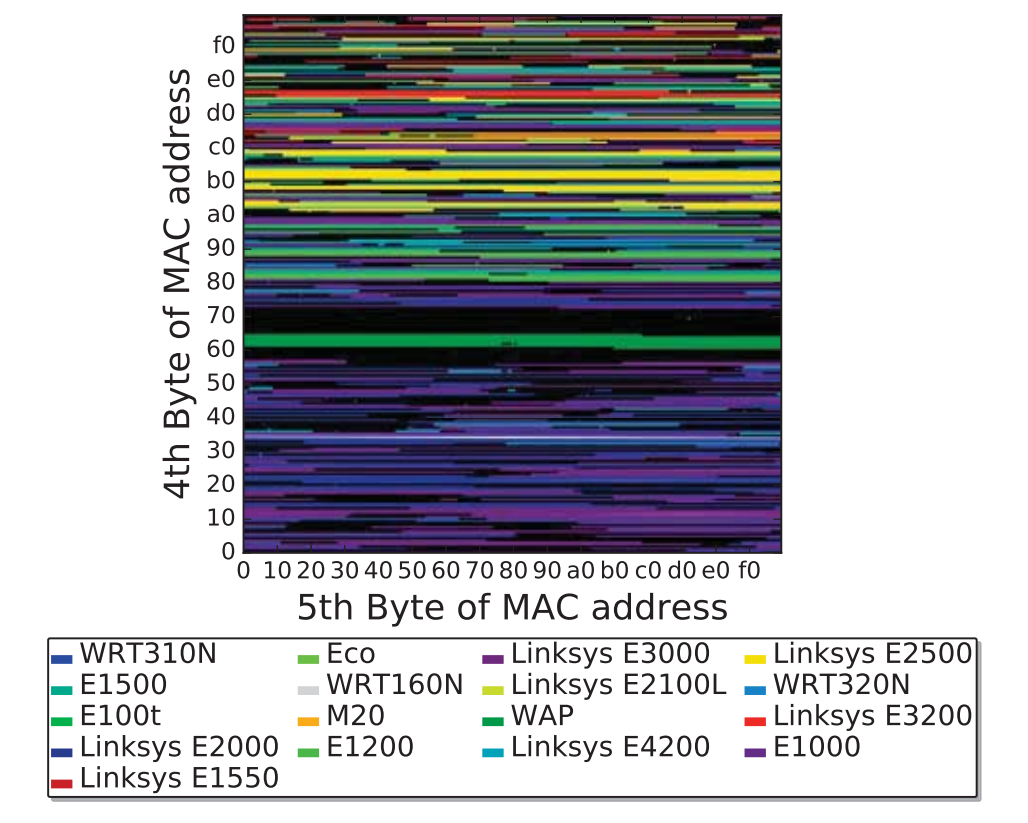}
        \caption{A Cisco OUI has 17 different models represented.}
        \label{fig:multi-cisco}
    \end{subfigure}%
    \caption{Two MAC address byte-axis plots demonstrating variation in MAC
    address assignment to models for two manufacturers.}
    \label{fig:multicolor}
\end{figure*}

If, as in Martin~\etal's ``Decomposition of MAC Address
Structure''~\cite{acsac2016furious} I am attempting to show both allocation
patterns and which belong to a certain categorical variable, like device model,
multiple hues may be used. Figure~\ref{fig:multicolor} displays  two examples of
a MAC address byte-axis plots that color ranges of MAC addresses by the model
they are associated with (in this case, model information was obtained using WPS
information elements from 802.11 probe requests and from multicast DNS.)
Figure~\ref{fig:multi-apple} represents four distinct Apple models; the iPhone
5c is allocated the majority of the OUI, though different models of iPad Mini 2s
also have nontrivial representation. In Figure~\ref{fig:multi-cisco}, much more
granular allocations are assigned to 17 different models. This is potentially a
result of Cisco acquiring OUI space previously assigned to companies like
Linksys, which it then further allocated.

\subsection{\vsix Allocation Byte-Axis Plots}

Like MAC address byte-axis plots, \vsix byte-axis plots also display address
allocations from a larger contiguous region of address space. In ``Follow the
Scent: Defeating IPv6 Prefix Rotation Privacy''~\cite{rye2021follow}, I used
\vsix byte-axis plots to display which /64 networks were assigned to customers
from an ISP's /48.

\begin{figure*}[t]
\centering
    \begin{subfigure}[c]{.48\textwidth}
        \includegraphics[width=\linewidth]{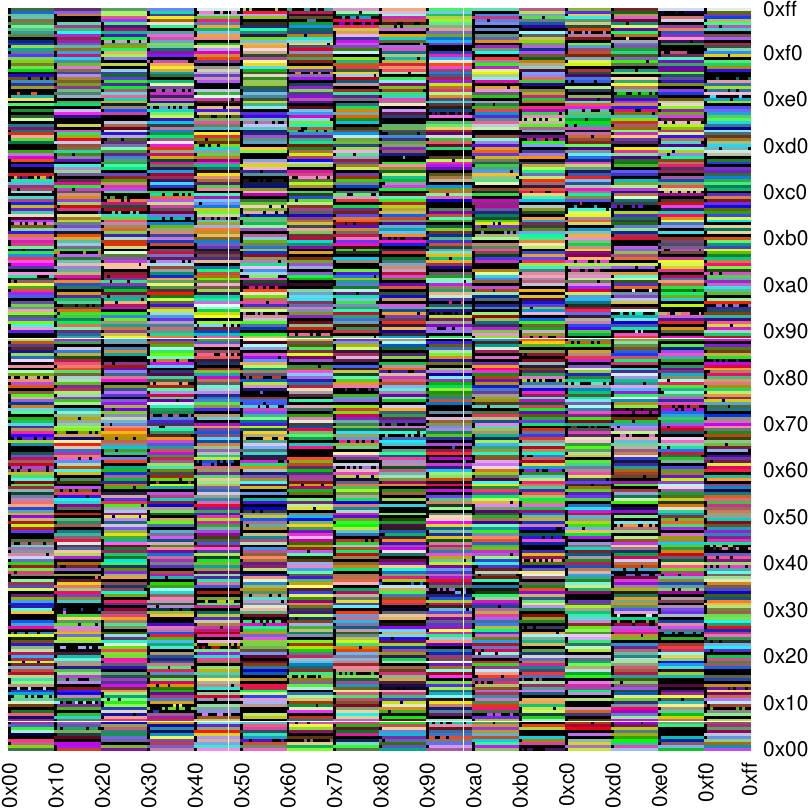}
        \caption{A byte-axis plot for a BH Telecom (Bosnia) prefix
        (\texttt{2a02:27b0:4a01::/48}.) }
        \label{fig:bihnet}
    \end{subfigure}%
    \hfill
    \begin{subfigure}[c]{.48\textwidth}
        \includegraphics[width=\linewidth]{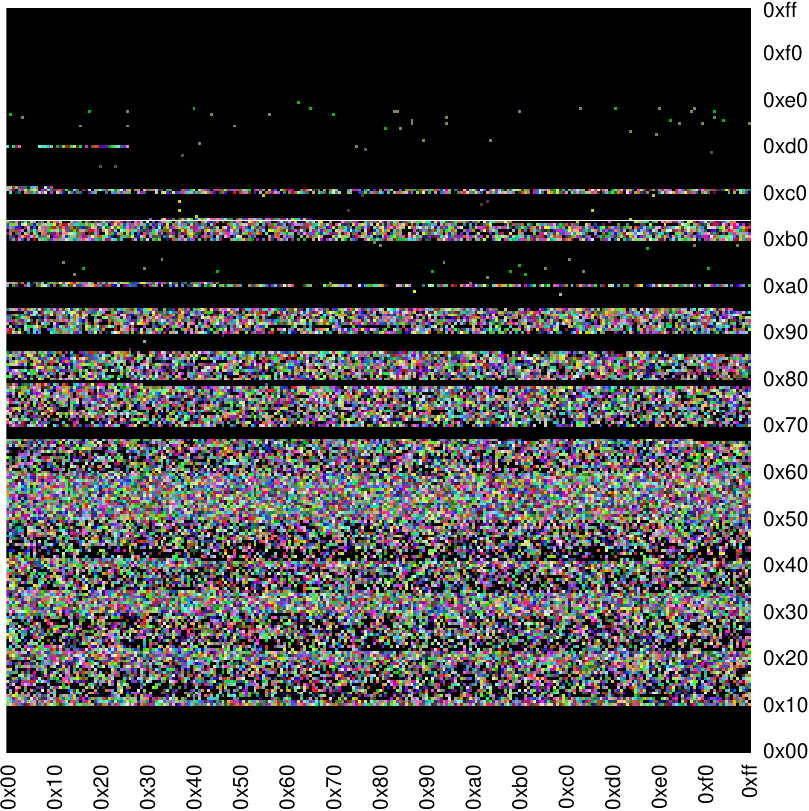}
        \caption{A byte-axis plot for a Starcat Japan /48 prefix
        (\texttt{2001:03b0:0022::/48})}
        \label{fig:mediacat}
    \end{subfigure}%
    \caption{Two IPv6 byte-axis plots of differently-sized allocation schemes.}
    \label{fig:v6-plots}
\end{figure*}

\begin{figure*}[tb]
\centering
    \begin{subfigure}[c]{.48\textwidth}
        \includegraphics[width=\linewidth]{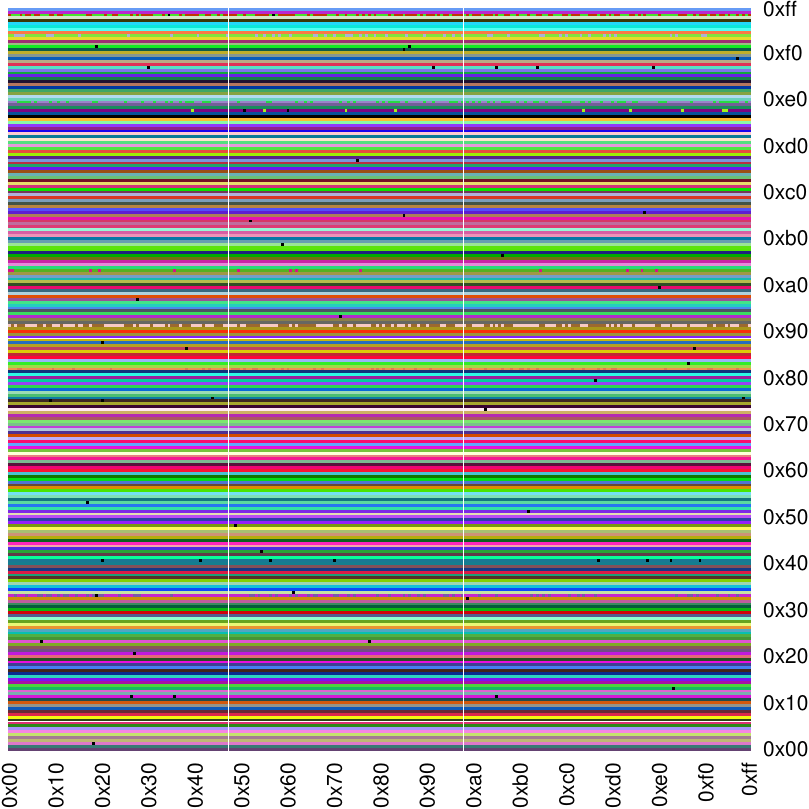}
        \caption{A Versatel 1\&1 /48 whose allocations are /56s.}
        \label{fig:versatel-56}
    \end{subfigure}%
    \hfill
    \begin{subfigure}[c]{.48\textwidth}
        \includegraphics[width=\linewidth]{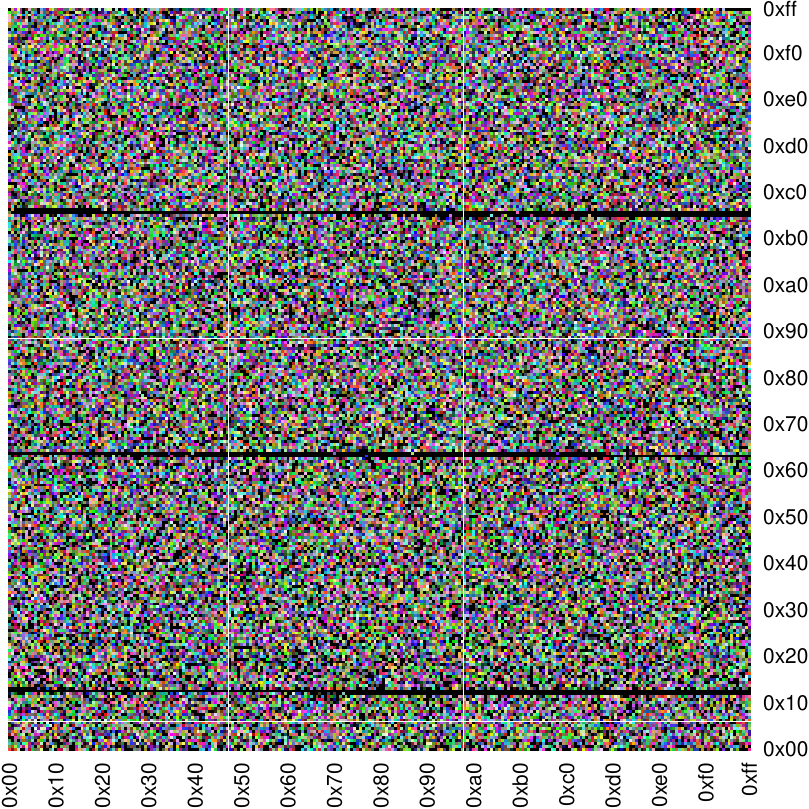}
        \caption{A Versatel 1\&1 /48 whose allocations are /64s.}
        \label{fig:versatel-64}
    \end{subfigure}%
    \caption{Two Versatel 1\&1 /48s with different allocation schemes.}
    \label{fig:two-versatel}
\end{figure*}

Figure~\ref{fig:v6-plots} displays two of those plots. Each subfigure represents
a different /48 from a residential ISP's address space. In
Figure~\ref{fig:bihnet}, a Bosnian ISP BH Telecom /48 is depicted. The $y$-axis
represents the $7^{th}$ byte of addresses within the /48, while the $x$-axis
represents the $8^{th}$. Each point in the plot, therefore, represents a /64
network (\texttt{2a02:27b0:4a01:yyxx}.) Individual colors represent responses
from the same IP address for contiguous
allocations of address space, although the hue itself was chosen randomly
and has no meaning. Coloring address allocations in this manner allows us to easily
understand that this /48 is subdivided into /60 allocations, because each
horizontal bar of the same color in the plot is 16 points wide and because each
color represents the same responsive IPv6 address (presumably a home router.)

In contrast, Figure~\ref{fig:mediacat} displays a /48 from Starcat Japan's
address space. As with the MAC address byte-axis plots, I use black points to
indicate parts of the /48 address space for which I have no observations.
However, the colored points indicate that Starcat does not follow the same
address assignment practices as BH Telecom, instead assinging /64s to its
customers. These appear as individually-colored points on the plot, with each
distinct hue representing a different responsive address.

In addition to illuminating \vsix allocation strategies between different
service providers, \vsix byte-axis plots can even demonstrate differences in
assignment policy within the same service provider.

Figure~\ref{fig:two-versatel} displays two Versatel 1\&1, a German consumer ISP,
/48s with different customer allocation strategies. Figure~\ref{fig:versatel-56}
displays a Versatel /48 that is subnetted into /56s which are allocated to their
customers, as evidenced by the differently-colored horizontal banding pattern.
In contrast, Figure~\ref{fig:versatel-64} is almost entirely allocated in /64
chunks to customers. This demonstrates that even within a single service
provider, several different allocation schemes may exist. This may manifest in
several ``tiers'' of service, or different types (e.g. individual versus
business) of customers.

\section{Conclusion}

In this work, I have described the construction of ``byte-axis'' plots.
Byte-axis plots take some fixed allocation size (e.g. a /48 \vsix network), and
separate the next two bytes within that allocation along the $y$ and $x$ axes.
These plots are useful because they enable the viewer to easily discern patterns
in how these address blocks are subdivided.

Byte-axis plots are useful in visualizing two types of networking addresses --
MAC addresses and \vsix addresses. 

In this work, I introduced two different MAC address allocation patterns between two different
wireless NIC manufacturers. Then, I demonstrated two examples of how we can visualize the
assignment of those allocations to individual device models.

Finally, I also introduced several examples of how different \vsix ISPs assign
differently-sized subnets to their customers. One, for instance, assigned /56
networks to subscribers, while another /60s, and a third /64s. Interestingly,
there are also examples of a single ISP supporting several different subnet
allocation sizes, potentially indicating multiple levels or types of service
that it offers.

\bibliographystyle{plain}
\bibliography{refs}

\end{document}